\documentclass[prb,preprint,showpacs,amsmath,amssymb,superscriptaddress]{revtex4}
\usepackage{epsfig}
\usepackage{graphicx}
\usepackage{amsmath,amssymb}
\usepackage{pdfpages}

\begin{document}

\title{Experimental observation of the optical spin-orbit torque}

\author{N.~Tesa\v{r}ov\'a}
\affiliation{Faculty of Mathematics and Physics, Charles University in Prague, Ke Karlovu 3, 121 16 Prague 2, Czech Republic}

\author{P.~N\v{e}mec}
\affiliation{Faculty of Mathematics and Physics, Charles University in Prague, Ke Karlovu 3, 121 16 Prague 2, Czech Republic}

\author{E.~Rozkotov\'a}
\affiliation{Faculty of Mathematics and Physics, Charles University in Prague, Ke Karlovu 3, 121 16 Prague 2, Czech Republic}

\author{J.~Zemen}
\affiliation{School of Physics and Astronomy, University of Nottingham, Nottingham NG7 2RD, United Kingdom}
\affiliation{Institute of Physics ASCR, v.v.i., Cukrovarnick\'a 10, 162 53 Praha 6, Czech Republic}

\author{F.~Troj\'anek}
\affiliation{Faculty of Mathematics and Physics, Charles University in Prague, Ke Karlovu 3, 121 16 Prague 2, Czech Republic}

\author{K.~Olejn{\'{i}}k}
\affiliation{Institute of Physics ASCR, v.v.i., Cukrovarnick\'a 10, 162 53 Praha 6, Czech Republic}

\author{V.~Nov{\'ak}}
\affiliation{Institute of Physics ASCR, v.v.i., Cukrovarnick\'a 10, 162 53 Praha 6, Czech Republic}

\author{P.~Mal\'y}
\affiliation{Faculty of Mathematics and Physics, Charles University in Prague, Ke Karlovu 3, 121 16 Prague 2, Czech Republic}

\author{T.~Jungwirth}
\affiliation{Institute of Physics ASCR, v.v.i., Cukrovarnick\'a 10, 162 53 Praha 6, Czech Republic}
\affiliation{School of Physics and Astronomy, University of Nottingham, Nottingham NG7 2RD, United Kingdom}

\date{\today}


\maketitle

{\bf
Spin polarized carriers electrically injected into a magnet from an external polarizer can exert a spin transfer torque (STT)\cite{Ralph:2008_a}  on the magnetization. The phenomenon belongs to the area of spintronics research focusing on manipulating magnetic moments by electric fields and is the basis of the emerging technologies for scalable magnetoresistive random access memories.\cite{Chappert:2007_a} In our previous work we have reported experimental observation\cite{Nemec:2012_a} of the optical counterpart of STT\cite{Rossier:2003_a,Nunez:2004_b} in which a circularly polarized pump laser pulse acts as the external polarizer, allowing to study and utilize the phenomenon on several orders of magnitude shorter timescales than in the electric current induced STT. Recently it has been theoretically proposed\cite{Manchon:2008_b,Manchon:2009_a,Garate:2009_a} and experimentally demonstrated\cite{Chernyshov:2009_a,Miron:2010_a,Fang:2010_a} that in the absence of an external polarizer, carriers in a magnet under applied electric field can develop a non-equilibrium spin polarization due to the relativistic spin-orbit coupling, resulting in a  current induced spin-orbit torque (SOT) acting on the magnetization.  In this paper we report the observation of the optical counterpart  of SOT.  At picosecond time-scales, we detect excitations of magnetization of a ferromagnetic semiconductor (Ga,Mn)As which are independent of the polarization of the pump laser pulses and are induced by non-equilibrium spin-orbit coupled photo-holes.
}

In current induced STT, spin-polarized carriers are electrically injected into a magnetic object, such as thin ferromagnetic layer or domain wall,  from another part of a non-uniform magnetic structure. The physical origin of STT is in the angular momentum transfer from the injected carrier spins to the magnetic moments. The current induced SOT, on the other hand, is observed in uniform magnets with no external source of spin polarized carriers. The non-equilibrium spin polarization of carriers producing SOT results from current induced redistribution of carrier states in the band structure of the magnet. The physical origin of SOT is the spin-orbit coupling in the carrier bands. While the seminal works on current induced STT are more than 15 years old\cite{Slonczewski:1996_a,Berger:1996_a} and the effect already plays a key role in commercially developed spintronic technologies, the research of the relativistic SOT is still at its infancy. Yet, the remarkable property of this inverse magneto-transport  effect, allowing a single piece of magnet to excite itself under applied electric field, has already found  practical utility. For example, when combined with the self-detection of the magnetization variations by anisotropic magnetoresistance, which is a direct magneto-transport effect based also on spin-orbit coupling, an all-electric ferromagnetic resonance measurement of micromagnetic parameters can be performed on a single ferromagnetic nanostructure.\cite{Fang:2010_a} 

The aim of our works reported in Ref.~\onlinecite{Nemec:2012_a} and in this paper is to find the light induced torque counterparts of STT and SOT. Ferromagnetic semiconductor (Ga,Mn)As utilized in our experiments is a favorable candidate material for observing the phenomena. The direct-gap GaAs host allows the generation of high density non-equilibrium photo-carriers and the carrier spins interact with ferromagnetic moments on Mn via strong exchange coupling.\cite{Jungwirth:2006_a} When the ferromagnetic Mn moments are excited, this can be sensitively detected by  probe laser pulses due to large magneto-optical (MO) signals in (Ga,Mn)As. Several groups have reported MO studies of fast laser induced magnetization dynamics in (Ga,Mn)As (see Supplementary information).\cite{Oiwa:2005_a,Wang:2006_b,Takechi:2007_a,Qi:2007_a,Qi:2009_a,Rozkotova:2008_a,Rozkotova:2008_b,Hashimoto:2008_a,Hashimoto:2008_b,Kobayashi:2010} However, the direct search and observation of the light induced torque counterparts of STT and SOT have not been the subject of these studies until our  work in Ref.~\onlinecite{Nemec:2012_a} and the work presented in this paper. 

In the optical spin transfer torque (OSTT), observed in our previous experiments in Ref.~\onlinecite{Nemec:2012_a}, the external source for injecting spin polarized carriers is provided by circularly polarized light at normal incidence which yields high degree of out-of-plane spin-polarization of injected photo-carriers due to the optical selection rules in GaAs. For the photo-electrons producing the OSTT, the spin-lifetime is much larger than the precession time in the exchange field of the Mn moments. This results in the steady state carrier spin density component oriented in the plane of the ferromagnetic film and perpendicular to the magnetization vector, which is an analogous mechanism to the adiabatic current-induced STT.\cite{Nemec:2012_a} Since large OSTT requires large spin lifetime of injected carriers, i.e. spin-orbit coupling is detrimental for OSTT, the weakly spin-orbit coupled photo-electrons play the key role in this case.\cite{Rossier:2003_a,Nunez:2004_b,Nemec:2012_a} 

In the optical spin-orbit torque (OSOT), the absence of an external polarizer corresponds to photo-carrier excitations which are independent of the polarization of the pump laser pulses.  Since the phenomenon relies on spin-orbit coupling, the non-equilibrium photo-holes are essential in this case. Because of the lack of external polarizer and because of strong spin-orbit coupling, precession of carriers in the exchange field of magnetic moments is not the origin of the torque in this case, in close analogy to the current induced SOT and reminiscent of the non-adiabatic STT.\cite{Manchon:2008_b,Manchon:2009_a,Garate:2009_a,Nemec:2012_a} The physical picture of the OSOT is as follows: The optically injected photo-holes relax towards the hole Fermi energy of the p-type (Ga,Mn)As on a short ($\sim100$~fs) timescale, creating a non-equilibrium excess hole density in the valence band. The change in the occupation of the hole states, as compared to the equilibrium state in dark, and the spin-orbit coupling can generate a non-equilibrium   hole spin polarization which is misaligned with the equilibrium orientation of Mn moments. This non-equilibrium photo-hole polarization  persists over the timescale of the hole recombination ($\sim$ps) during which it exerts a torque on the Mn local moments via the kinetic-exchange coupling. The three key characteristic features, i.e., the non-equilibrium occupation of carrier states, spin-orbit coupling in carrier bands, and the resulting non-equilibrium carrier polarization acting on the magnetic moments via exchange coupling makes OSOT the direct optical counterpart of the current induced SOT. Since the applied electrical drift (and relaxation) yields a non-equilibrium carrier occupation in the form of asymmetric redistribution on the Fermi surface, the current induced SOT is observed in systems with broken inversion symmetry in the crystal.  OSOT is caused by optical generation (and relaxation) of photo-carriers  without an applied drift and, therefore, the broken inversion symmetry in the crystal is not required. It is  replaced by the broken time reversal symmetry, i.e., by a non-zero spin polarization of  carrier bands in equilibrium. This is another reason why holes in the strongly exchange-split valence band in (Ga,Mn)As govern OSOT.

The schematic illustration and our experimental observation of OSOT are shown in Fig.~1. Apart from  the physically more intriguing nature of relativistic OSOT, as compared to OSTT, its experimental identification is complicated by the presence of thermal excitation mechanisms of magnetization dynamics in the case of pump-polarization independent signals (see Refs.~\onlinecite{Wang:2006_d,Kirilyuk:2010_a} and Supplementary information). The absorption of the pump laser pulse leads to photo-injection of electron-hole pairs. The non-radiative recombination of photo-electrons produces a  transient increase of lattice temperature which builds up on the time scale of $\sim 10$~ps and persists over $\sim1000$~ps. This results in a quasi-equilibrium easy-axis (EA) orientation which is tilted from equilibrium EA.  Consequently, Mn moments in (Ga,Mn)As start to precess around quasi-equilibrium EA, as illustrated in Fig.~1a, with a typical precession time of $\sim 100$~ps given by the magnetic anisotropy fields in (Ga,Mn)As. As discussed in detail below, EA stays in-plane and the sense of rotation within the plane of the (Ga,Mn)As film with increasing temperature is uniquely defined. In the notation introduced in Fig.~1c, the change of the in-plane angle $\delta\varphi$ of magnetization during the thermally excited precession  can be only positive. OSOT, illustrated in Fig.~1b, acts during the laser pulse (with a duration of 200~fs) and fades away within the hole recombination time ($\sim$~ps). It causes an impulse tilt of the magnetization which allows us to clearly distinguish OSOT from the considerably slower thermal excitation mechanism. Moreover, the initial OSOT induced tilt of magnetization can yield precession angles that are inaccessible in the thermally induced magnetization dynamics. This provides another evidence for OSOT. 
\begin{figure}[h!]
\includegraphics[width=.7\columnwidth,angle=0]{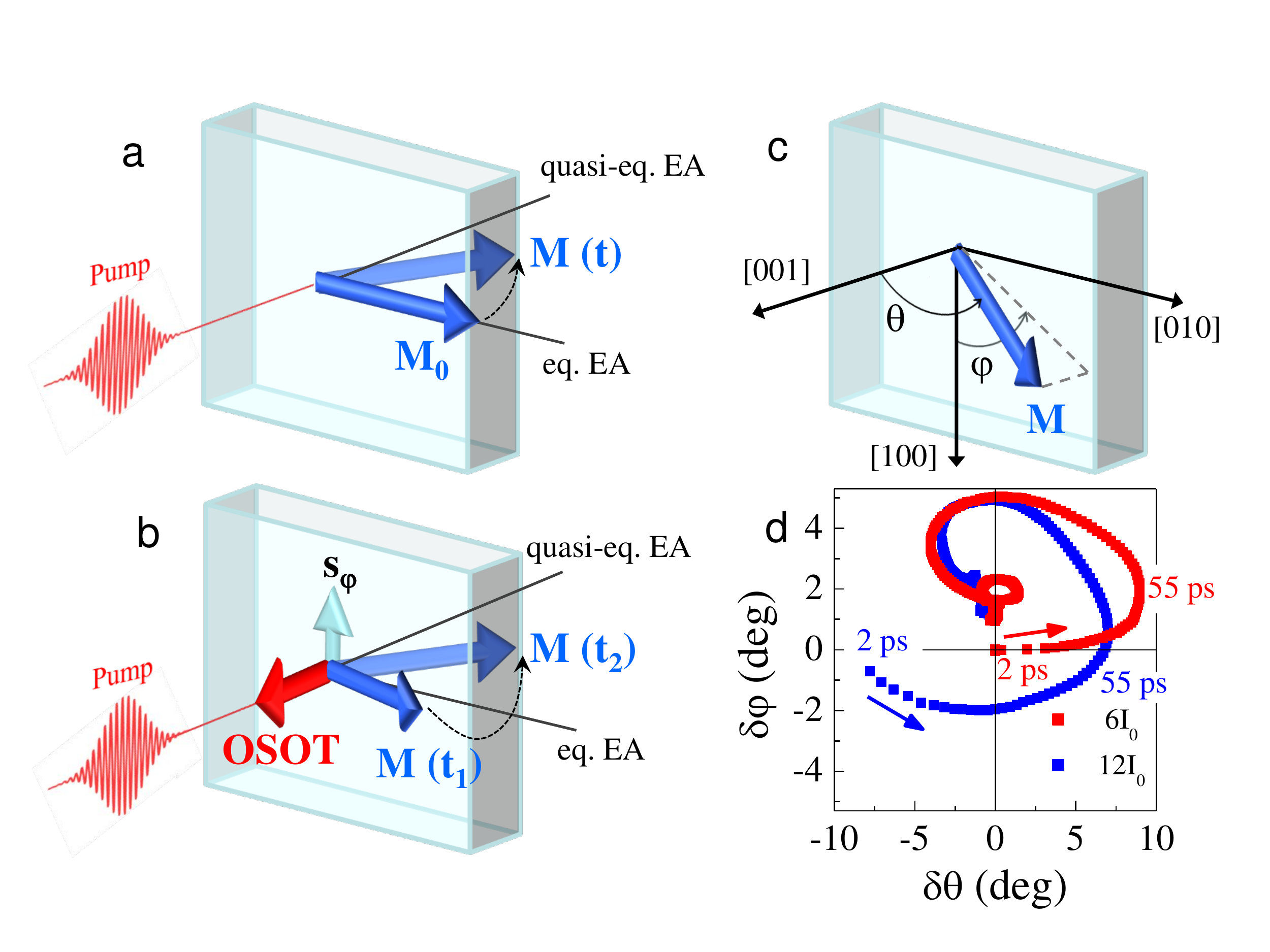}
\caption{{\bf Schematic illustration and experimental observation of the optical spin-orbit torque.} {\bf a,} Schematic illustration of the thermally excited precession of magnetization {\bf M}(t) around the transient quasi-equilibrium easy axis (EA). {\bf M}$_0$ is the magnetization vector aligned with  in-plane equilibrium EA before the pump pulse.  {\bf b,} Schematic illustration of the OSOT induced by the in-plane transverse component {\bf s}$_{\varphi}$ of the non-equilibrium hole spin polarization. On the time-scale of magnetization precession, OSOT causes an instantaneous tilt of the magnetization {\bf M}(t$_1$) which allows us to clearly distinguish OSOT from the considerably slower thermal excitation mechanism. The initial OSOT induced tilt of magnetization can yield precession angles that are inaccessible in the thermally induced magnetization dynamics. {\bf c,} Definition of the coordinate system. {\bf d,} Time evolution of the magnetization vector measured  in a (Ga,Mn)As material with nominal Mn-doping $x=3$\%. The direction of the time increase is depicted by arrows. Magnetization tilt angles $\delta\varphi$ and $\delta\theta$ are measured with respect to  equilibrium EA. Sample base temperature before pump pulse was 15 K and experiments were performed at zero magnetic field. At lower pump intensity $6I_0$ ($I_0 = 7$~$\mu$Jcm$^{-2}$) the precession is induced thermally while at $12I_0$ the OSOT induced initial tilt is observed. 
}
\label{fig1}
\end{figure}

Examples of the direct observation of the thermally governed excitation of magnetization at a lower pump pulse intensity of $6I_0$ and of the OSOT governed excitation at a higher intensity of $12I_0$ are shown in Fig.~1d, where $I_0=7$~$\mu$Jcm$^2$. We point out that these dynamical MO signals are completely independent of the polarization of pump pulses, i.e., they are the same for any orientation of the polarization plane of linearly polarized  pump laser pulses and they also correspond to the polarization-independent part of the MO signals extracted by summing the signals induced by $\sigma^+$ and $\sigma^-$ circularly polarized pump pulses (see Supplementary information). The distinct features of OSOT, described in the previous paragraph, are clearly visible when comparing the two measured  trajectories of  magnetization angles. This key demonstration  has been enabled by the technique which we developed in Ref.~\onlinecite{Tesarova:2012_a} and which translates the measured dynamical MO signals in our pump-and-probe experiments to the time-dependent magnetization vector trajectory. This is done without assuming any theoretical model for the magnetization dynamics and without using any fitting parameter. Our experimental method utilizes different dependences of the polar Kerr effect (PKE) and magnetic linear dichroism (MLD) on the orientation of linear polarization of the probe laser pulses to disentangle the contributions to the MO signal from the out-of-plane and in-plane components of the magnetization motion, respectively. The magnitudes of PKE and MLD coefficients in a particular (Ga,Mn)As sample are determined from static MO experiments in which an external magnetic field is used to align the magnetization in a defined orientation (see Supplementary information). To obtain each point  on the trajectory of the  magnetization vector excited in the pump-and-probe experiment, we performed a set of measurements of the MO signal as a function of  the orientation of the polarization plane of the linearly polarized probe pulse. The dynamical MO measurements were performed at zero magnetic field; prior to the experiment, the magnetization was aligned with EA. (For more details on the experimental technique see Refs.~\onlinecite{Nemec:2012_a,Tesarova:2012_a} and Supplementary information.)   

Our observation of OSOT stems from an extensive growth, characterization, and MO study of a series of (Ga,Mn)As/GaAs materials. The ferromagnetic semiconductor epilayers have individually optimized molecular-beam-epitaxy and post-growth annealing procedures for each nominal Mn doping in order to minimize the density of compensating defects and other unintentional impurities and to achieve high uniformity of the epilayers. Nominal Mn-dopings in this set of 20~nm thick ferromagnetic (Ga,Mn)As epilayers span the whole range from 1.5\% up to $\sim 14$\% (i.e. up to $\sim 8$\% of uncompensated Mn$_{\rm Ga}$ impurities) with ferromagnetic transition temperatures reaching 188~K.  All samples within the series have reproducible characteristics with the overall trend of increasing Curie temperature, increasing hole concentration, and increasing magnetic moment density with increasing nominal Mn doping.\cite{Jungwirth:2010_b}  The samples are in-plane magnets in which the biaxial anisotropy, reflecting the cubic symmetry of the host crystal, competes with an additional uniaxial anisotropy. The biaxial anisotropy dominates at very low dopings and the easy axis aligns with the main crystal axis [100] or [010]. At intermediate dopings, the uniaxial anisotropy is still weaker but comparable in magnitude to the biaxial anisotropy. In these samples the two equilibrium easy-axes are tilted towards the [$\bar1$10] direction and are sensitive to small changes in doping or temperature. At very high dopings, the uniaxial anisotropy dominates and the system has one strong easy-axis along the [$\bar1$10] in-plane diagonal.

In Fig.~2 we show examples of measured dynamical MO signals in several samples with different magnetic anisotropies. In agreement with the doping trends in static magnetic anisotropies described above, we do not observe laser induced precession of magnetization in the very low and very high doped samples in which the magnetic easy axis aligns with one of the high symmetry crystal direction and its direction is insensitive to small perturbations. On the other hand, we observe precessions in samples with intermediate doping in which the EA direction is sensitive to the sample temperature and the hole concentration. Note that the observed precession frequencies in the studied set of samples correspond to magneto-crystalline anisotropy fields which are fully consistent with the respective anisotropy fields obtained from magnetization measurements by the superconducting quantum interference device (SQUID). 

\begin{figure}[h!]
\includegraphics[width=0.7\columnwidth,angle=0]{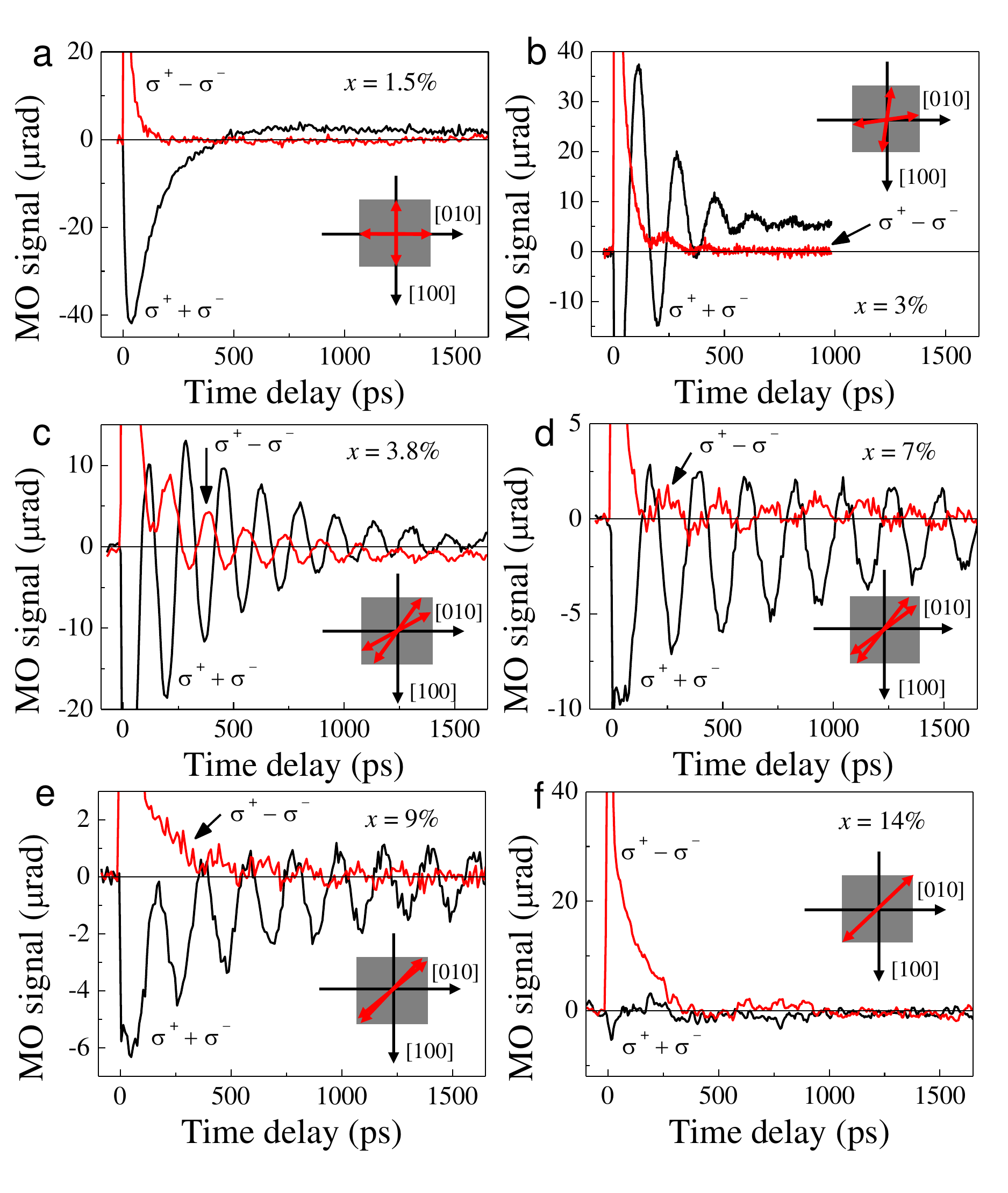}
\caption{{\bf Laser pulse-induced precession of magnetization in a series of optimized  (Ga,Mn)As epilayers.} {\bf a-f,} Dynamical MO signals in samples with nominal Mn-doping $x=1.5-14$\% induced by circularly ($\sigma^+$ and $\sigma^-$) polarized pump pulses were measured from which the helicity-sensitive [($\sigma^+-\sigma^-$)/2] and polarization-insensitive  [($\sigma^++\sigma^-$)/2] parts of the signals were extracted. Pump intensity in these measurements was $4I_0$, sample base temperature before pump pulse was 15 K, and external magnetic field of 20 mT was applied along the [010] crystal axis. Insets illustrate variations of easy axis positions across the series of measured (Ga,Mn)As samples at 15~K.}
\label{Fig2}
\end{figure}

The MO signals which depend on the helicity of circularly polarized pump laser pulses, labeled in Fig.~2 as $\sigma^+-\sigma^-$, originate from the excitation of ferromagnetic Mn moments by OSTT. The key signatures of the OSTT induced MO signals are extensively discussed  in Ref.~\onlinecite{Nemec:2012_a}. The aim of  the present work is the observation of OSOT and we therefore focus on magnetization precession signals which do not depend on the polarization of pump laser pulses. The corresponding data are labeled as $\sigma^++\sigma^-$ in Fig.~2. Since the misalignment of the non-equilibrium hole polarization with the equilibrium orientation of Mn moments  has the same physical origin as the dependence of the static magnetic easy-axis orientation in (Ga,Mn)As materials on hole density, the observation of OSOT requires a (Ga,Mn)As film which shows sensitivity of the easy-axis orientation to small changes in doping.\cite{Zemen:2009_a} For measurements shown in Fig.~1d and below in Figs.~3,4  we have singled out a material from the lower doping end in our series. This 3\% Mn doped epilayer (with a Curie temperature of 77~K) is still a relatively low hole-density material but with already competing biaxial and uniaxial anisotropies for which we can expect sizable OSOT at accessible laser intensities. 

Detailed measurements of the pump-polarization independent magnetization trajectories for several intensities of the pump pulses are shown in Fig.~3. The key characteristics of the magnetization dynamics at low intensities $I_0$ and $6I_0$ reflect the thermal excitation mechanism described in Fig.~1a. In equilibrium, EA in the 3\% Mn doped sample is tilted by approximately 10$^\circ$ from the [010] ($\varphi=90^\circ$) crystal axis towards the  [$\bar1$10] ($\varphi=135^\circ$) in-plane diagonal direction. With increasing temperature, the easy-axis rotates further towards the [$\bar1$10] direction. This is because the uniaxial anisotropy component scales with magnetization as $\sim M^2$ while the biaxial component scales as $\sim M^4$ and, therefore, the uniaxial anisotropy gets enhanced relative to the biaxial anisotropy with increasing temperatures. This expected EA rotation is confirmed by independent SQUID measurements and microscopic calculations based on the ${\bf k}\cdot {\bf p}$ kinetic-exchange Hamiltonian,\cite{Jungwirth:2006_a,Zemen:2009_a} shown in Fig.~4a,b. The amplitude of the precession angles for the intensity $6I_0$ (Fig.~3b) is larger than for the intensity $I_0$ (Fig.~3a). This is consistent with a larger increase of the transient temperature (and corresponding larger tilt of quasi-equilibrium EA) for the larger pump pulse intensity. We have deduced the temperature increase due to pump pulses from the measured precession frequencies which reflect the temperature dependent magnetocrystalline anisotropy energies.  In Fig.~4c we plot the dependence of the precession frequency on the base sample temperature at low excitation intensity $I_0$ and on the laser intensity at low base temperature of 15~K. From the comparison of these two measurements we infer the magnitude of the transient temperature change as a function of the laser intensity. (Note that consistent temperature vs. intensity calibration is obtained from the comparison of the intensity dependence of the pump-induced demagnetization and the temperature dependence of the remanent magnetization measured by SQUID.) Fig.~4c confirms a sizable difference in transient temperatures for intensities $I_0$ and $6I_0$.

\begin{figure}[h!]
\includegraphics[width=.7\columnwidth,angle=0]{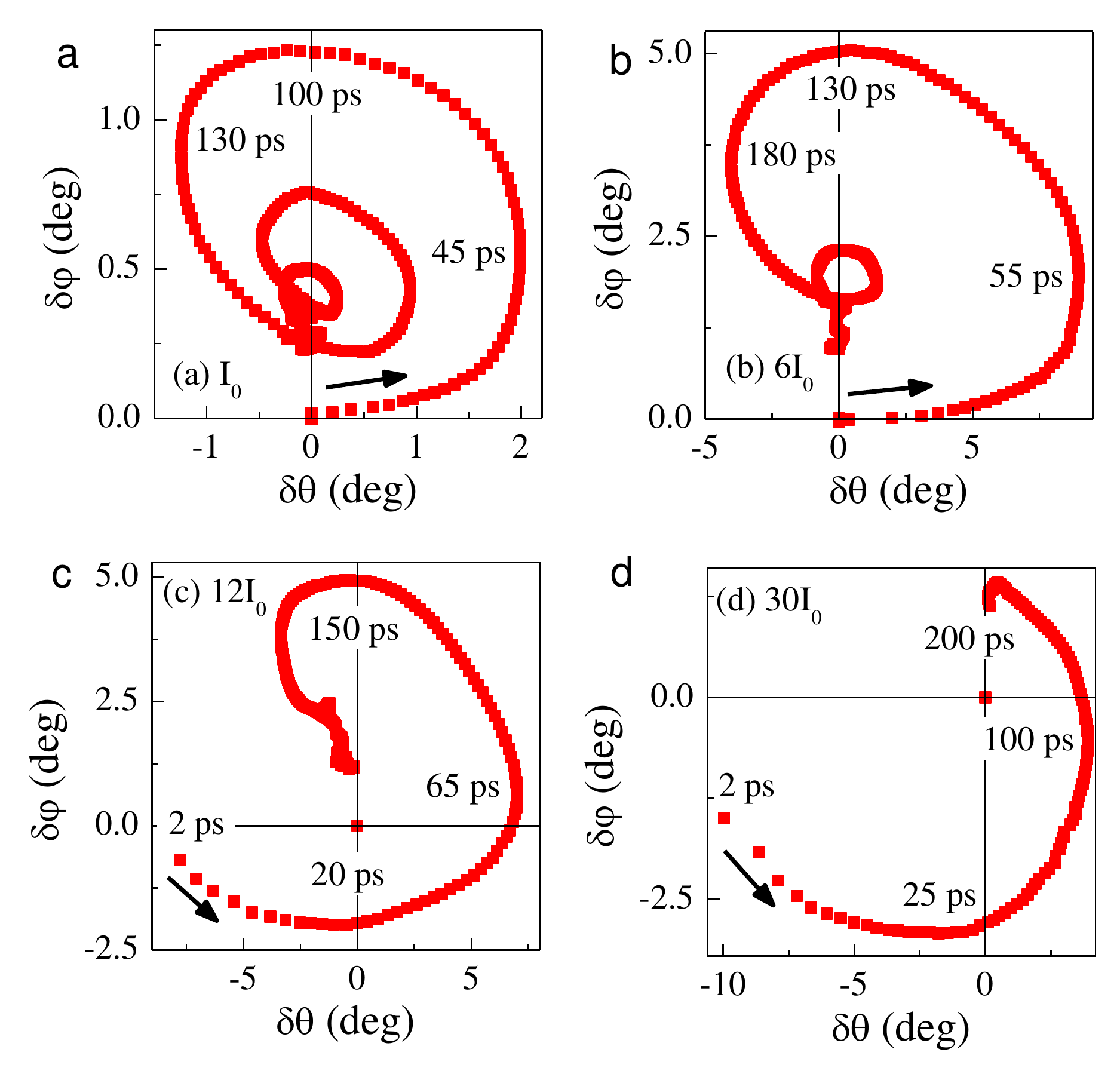}
\caption{{\bf Direct experimental reconstruction of the magnetization  trajectory from MO signals.} {\bf a-d,} Time evolution of the magnetization vector measured in the 3\% Mn-doped (Ga,Mn)As at pump intensities $I_0-30I_0$. The plotted data correspond to time delays between pump and probe pulses in steps of 2~ps. Sample base temperature before pump pulse was 15 K and experiments were performed at zero magnetic field. 
}
\label{Fig3}
\end{figure}

Remarkably, the heating of the sample by the laser pulses saturates at approximately $10 I_0$, as seen from Fig.~4c. The measured trajectories of the dynamical magnetization vector, however, show dramatic differences below and above $10 I_0$. The impulse tilt and precession angles inaccessible by the thermal excitations, seen in Figs.~3c,d for intensities  $12 I_0$ and $30 I_0$, were already pointed out in Fig.~1 as key signatures of OSOT. The complete saturation of the transient temperature increase  at $10 I_0$ provides another confirmation that the magnetization dynamics at high pump pulse intensities are governed by a distinct non-thermal mechanism. The connection between OSOT and photo-carrier generation is evidenced in Fig.~4d. Here we show the observed  change in the measured reflectivity of the (Ga,Mn)As film which correlates with the number of generated photo-carriers (see Supplementary information). The pump-induced change of the index of refraction is linear in pump-pulse intensity up to $\approx 25 I_0$ after which it starts to  saturate. It means that, unlike the transient temperature, the number of generated photo-carriers keeps increasing with increasing pump pulse intensity above  $10 I_0$. The concentration of photo-injected carriers can be estimated from the laser spot size, photon energy, and absorption and reflection coefficients of the sample. For the higher intensity measurements, the obtained photo-carrier density is of the order of $\sim 10^{19}-10^{20}$~cm$^{-3}$.

\begin{figure}[h!]
\includegraphics[width=.7\columnwidth,angle=0]{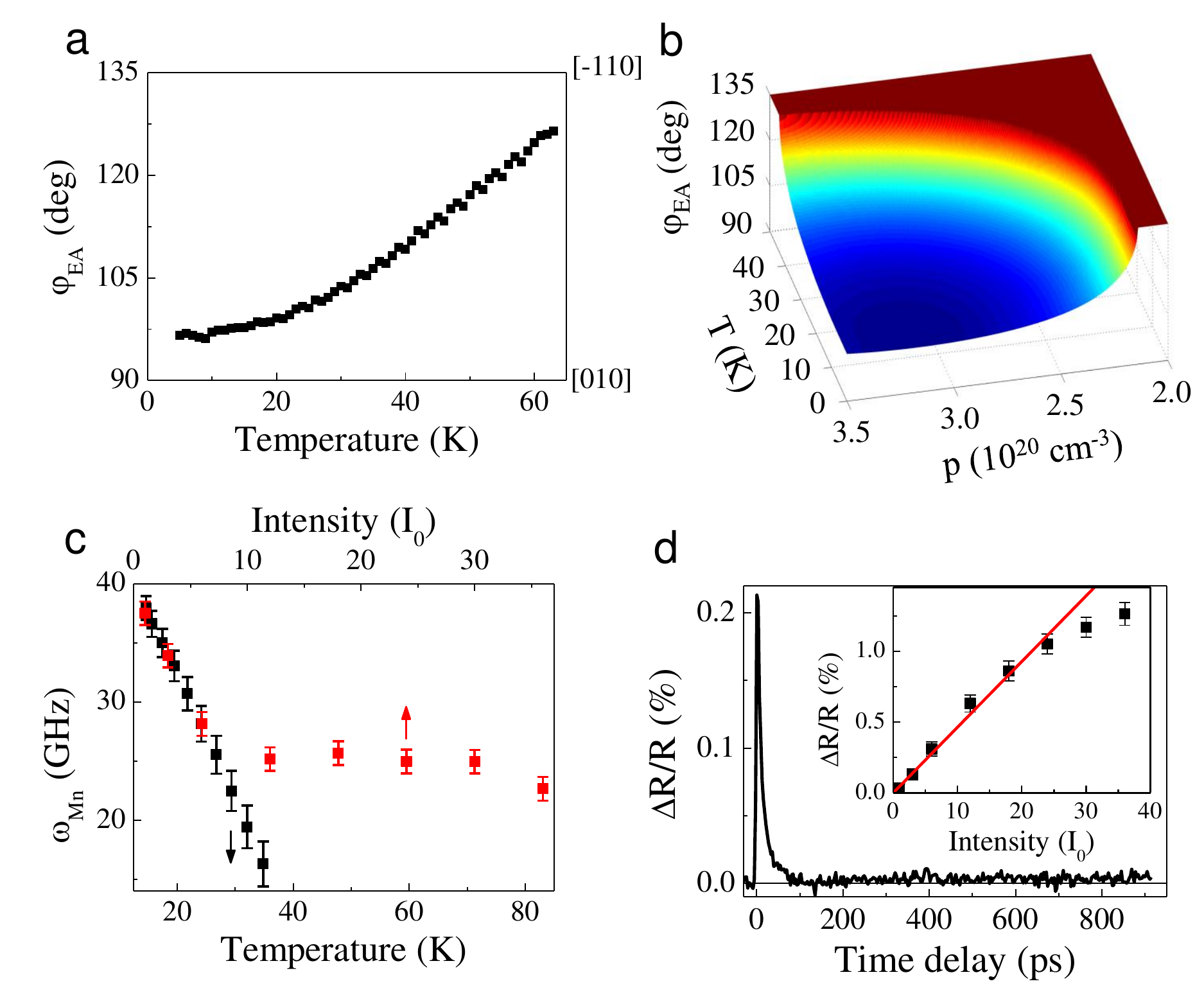}
\caption{{\bf Characterization of the 3\% Mn-doped (Ga,Mn)As.} {\bf a,} Temperature dependence of the equilibrium EA orientation determined from SQUID magnetization measurements. Consistent EA orientations are inferred from the MO experiments. {\bf b,} Microscopic calculations of the temperature and hole density dependent EA orientation. {\bf c,} Frequency of precessing Mn moments measured at zero magnetic field as a function of the base temperature at low excitation intensity $I_0$, and as a function of the  pump intensity at low base temperature of 15 K. {\bf d,} Dynamics of pump-induced reflectivity change at intensity $6I_0$. Inset shows the intensity dependence of the initial reflectivity change at base temperature of 15 K (line depicts a linear dependence).
}
\label{Fig4}
\end{figure}

In Fig.~4b we show that  equilibrium EA is sensitive to the hole density variations and that the sense of the tilt of EA with increasing hole density can be opposite than in the case of the temperature increase. Although not explaining the experiments on a quantitative level, Fig.~4b provides the clue why the instantaneous out-of-plane component of the tilt of the magnetization due to OSOT at higher pump intensities is opposite than the initial out-of-plane component of the precessing magnetization around  thermally excited quasi-equilibrium EA at lower pump intensities. A full quantitative theory of OSOT is a challenging problem compared to the theory of OSTT.\cite{Rossier:2003_a,Nunez:2004_b,Nemec:2012_a} In the latter case, the non-equilibrium spin-density of weakly spin-orbit coupled photo-electrons, producing the OSTT, is directly determined by the external polarizer, i.e., by the intensity, propagation axis, and helicity of the circularly polarized pump laser beam. The relation between the non-equilibrium density of photo-holes and the transverse component of their spin polarization, producing OSOT  in the case of light-polarization independent excitations, results from a more complex interplay of  spin-orbit coupling and photo-generation and relaxation processes in the spin-split valence band of the ferromagnetic semiconductor.  The effective field $J{\bf s}$ generating OSOT,
\begin{equation}
\frac{d\hat{\bf M}}{dt}=\frac{J}{\hbar}\hat{\bf M}\times{\bf s}\,,
\label{OSOT}
\end{equation}
where ${\bf s}$ is the non-equilibrium hole spin polarization density and $J\approx 50$~meVnm$^3$  is the hole - Mn moment exchange coupling constant, can be related in a simplified picture to the hole-density dependent magnetocrystalline anisotropy field (see Supplementary information for more theory details).  Using again the ${\bf k}\cdot {\bf p}$ kinetic-exchange Hamiltonian $H=H_{host}+JN_{Mn}S{{\bf \hat{M}}\cdot{\boldsymbol\sigma}}$,\cite{Jungwirth:2006_a,Zemen:2009_a} where $H_{host}$ is the host semiconductor Hamiltonian, $N_{Mn}$ is the Mn local moment density, $S=5/2$ is the local moment spin, ${\bf \hat{M}}$ is the local moment unit vector, and ${\boldsymbol\sigma}$ is the hole spin operator, the anisotropy field can be written as,\cite{Garate:2009_a}
\begin{eqnarray}
{\bf H}_{an}&=&-\frac{1}{N_{Mn}S}\frac{\partial}{\partial{\bf \hat{M}}}\sum_{a}\int d{\bf k}\epsilon_{a,{\bf k}}f_{a,{\bf k}}
=-\frac{1}{N_{Mn}S}\sum_{a}\int d{\bf k}\langle a,{\bf k}|\frac{\partial H}{\partial {\bf \hat{M}}}|a,{\bf k}\rangle f_{a,{\bf k}}\nonumber\\&=&-
\sum_{a}\int d{\bf k}\langle a,{\bf k}|J{\boldsymbol\sigma}|a,{\bf k}\rangle f_{a,{\bf k}}=-J{\bf s}\,.
\end{eqnarray}
Here $\epsilon_{a,{\bf k}}$ and $f_{a,{\bf k}}$ are the eigenenergy of $H$ and Fermi distribution function, respectively,  labeled by the band and wavevector index. Consistent with the in-plane orientation of  EA, we obtain that the out-of-plane transverse component of the anisotropy field, ${\bf H}_{an,\theta}=-J{\bf s}_{\theta}=0$ for any in-plane ${\bf \hat{M}}$ and any considered hole density. The in-plane transverse component, ${\bf H}_{an,\varphi}=-J{\bf s}_{\varphi}$, is zero when ${\bf \hat{M}}$ is aligned with  EA at a given hole density and non-zero for other orientations of ${\bf \hat{M}}$ at the same hole density. Since the EA orientation is sensitive to the hole density, as shown in Fig.~4b, ${\bf H}_{an,\varphi}$ for a given orientation of ${\bf \hat{M}}$ can change when the hole density is increased by, e.g., the photo-excitation (see Supplementary information).  The sign of the calculated ${\bf H}_{an,\varphi}$ is consistent with the sense of the initial magnetization tilt observed in experiments governed by the OSOT. The amplitudes of $J{\bf s}_{\varphi}$, obtained from the calculated hole density dependent anisotropy fields, are  $\sim$~$\mu$eV which is about 10 times smaller than the experimental strength of OSOT fields inferred form the measured out-of-plane tilts of the magnetization. The ${\bf k}\cdot {\bf p}$ kinetic-exchange model is known to underestimate by an order of magnitude anisotropy fields in the lower doped ferromagnetic (Ga,Mn)As samples.\cite{Zemen:2009_a}  Considering this general limitation of the model Hamiltonian theory we can conclude that the calculations confirm our experimental observation of OSOT.

To conclude, following the experimental discovery of OSTT induced by circularly polarized laser pulses, we have reported in the present paper the experimental observation of OSOT. The presence of this intriguing phenomenon, in which non-equlibrium photo-carrier polarization and the corresponding torque on magnetization are generated by spin-orbit coupling in the absence of an external polarizer, is evidenced by our  direct experimental measurement of magnetization vector trajectories. OSOT which induces magnetization excitations  independent of the polarization of pump laser pulses is observed in our experiments on picosecond timescales which is orders of magnitude shorter than in the current induced spin-orbit torque experiments. 


\section*{Acknowledgment}
We acknowledge fruitful discussions with Allan H. MacDonald, Jairo Sinova, and Jorg Wunderlich,  and support  from the EU ERC Advanced Grant No. 268066 and FP7-215368 SemiSpinNet, from the EPSRC Grant No. EP/H029257/1, from the Grant Agency of the Czech Republic Grant No. 202/09/H041 and P204/12/0853, from the Charles University in Prague Grant No. SVV-2012-265306 and No. 443011, and from the Academy of Sciences of the Czech Republic Preamium Academiae.

\protect\newpage

\vspace*{1cm}


\section*{Supplementary material}

# Experimental observation of the optical spin-orbit torque: Supplementary information

N. Tesařová,[1] P. Němec,[1] E. Rozkotová,[1] J. Zemen,[2,3] F. Trojánek,[1] K. Olejník,[3] V. Novák,[3] P. Malý,[1] and T. Jungwirth[3,2]

[1] *Charles University in Prague, Faculty of Mathematics and Physics, Ke Karlovu 3, 121 16 Prague 2, Czech Republic*

[2] *School of Physics and Astronomy, University of Nottingham, Nottingham NG72RD, United Kingdom*

[3] *Institute of Physics ASCR, v.v.i., Cukrovarnická 10, 16253 Praha 6, Czech Republic*

PACS numbers: 75.50.Pp, 76.50.+g, 78.20.Ls, 78.47.-p

**EXPERIMENTS PRECEEDING OUR OBSERVATIONS OF OSTT AND OSOT**

(Ga,Mn)As and related ferromagnetic semiconductors are potentially ideal test-bed materials for exploring laser induced excitations of magnetization. Their direct-gap band structure allows for strong optical excitations of the electronic system, the photo-carriers can directly interact with magnetic moments via strong exchange coupling, the carrier mediated ferromagnetism produces large and tunable magnetic and magneto-optical effects, and the relatively simple band structure is favorable for identifying microscopic physical origins of the phenomena. Prior to our observations of optical spin transfer torque (OSTT) and optical spin-orbit torque (OSOT) presented in Ref. r11b and in this work, femtosecond laser pulse induced precession of magnetization in ferromagnetic (Ga,Mn)As has been reported by several groups [r1–r8]. In Ref. r3, magnetization precession was induced by laser pulses of a relatively weak intensity $\approx 1 - 10$ μJcm$^{-2}$ and ascribed to heating effects. On the other hand, in Refs. r1 and r7 the observed precession triggered by comparably weak laser intensities was attributed to an effect of photo-injected holes on the magnetic anisotropy. Apart from these competing views, the photo-carrier based interpretation cannot be reconciled with the theoretical and experimental understanding of static magneto-crystalline anisotropies in (Ga,Mn)As [r7]. The interpretation would imply sizable changes of magnetic anisotropy in materials with equilibrium hole densities $\approx 10^{20} - 10^{21}$ cm$^{-3}$ induced by photo-injected holes of density as low as $\approx 10^{16}$ cm$^{-3}$. To compare with, e.g., electrical gating experiments [r9,

r10], we recall that changes in the magnetic anisotropy are detected for field-induced hole accumulation or depletion of at least $\approx 10^{18} - 10^{19}$ cm$^{-3}$. We also point out that the original interpretation in Ref. r1 in terms of photo-carrier induced changes of in-plane anisotropy was subsequently revised by assuming [r7,r8] out-of-plane tilts of the easy-axis. Since out-of-plane easy-axis rotation is not consistent with measured trends in static magnetic anisotropies of the considered (Ga,Mn)As/GaAs materials this again illustrates that the non-thermal laser induced magnetization precession due to optically generated carriers in ferromagnetic semiconductors has remained in Refs. r1-r8 an attractive yet unproven concept. In Ref. 11b and in this work we provide both the direct experimental evidence for the non-thermal effects and the microscopic interpretation of the physical origin of the observed optical torques.

## SAMPLES

The time-resolved magneto-optical experiments described in the main text were performed in a large set of optimized (Ga,Mn)As epilayers. Ferromagnetic (III,Mn)V semiconductors, with (Ga,Mn)As as the most thoroughly investigated example, are materials grown under highly non-equilibrium conditions due to the extreme levels of Mn-doping. Therefore, a special care has to be taken when preparing the samples and when generalizing the experimental results obtained in a given sample to the universal behavior of this material system. Very recently we reported a systematic study of optical and magneto-optical properties of an optimized set of (Ga,Mn)As epilayers spanning the wide range of accessible substitutional Mn$_{Ga}$ dopings [r11]. The optimization of the materials in the series, which is performed individually for each nominal Mn doping, minimizes the uncertainties in the experimental sample parameters and produces high quality epilayers which are as close as possible to uniform uncompensated (Ga,Mn)As mixed crystals. For each nominal Mn doping $x$, the growth and post-growth annealing conditions were separately optimized in order to achieve the highest Curie temperature $T_c$ attainable at the particular $x$. The highest $T_c$ criterion was found to lead simultaneously to layers with maximized uniformity and minimized compensation by unintentional impurities and defects [r11]. All samples are in-plane magnets in which the cubic anisotropy competes with an additional uniaxial anisotropy. At very low dopings, the cubic anisotropy dominates and the easy axis aligns with the main crystal axis [100] or [010]. At intermediate dopings, the uniaxial anisotropy is still weaker but comparable in magnitude to the cubic anisotropy. At very high dopings, the uniaxial anisotropy dominates and the system has one strong easy-axis along the [1-10] in-plane diagonal. We confirmed

that the laser pulse-induced precession of magnetization can be observed in all (Ga,Mn)As epilayers except in those with very low and very high doping levels where one of the anisotropies strongly dominates - see Table I.

| sample | $x$ (%) | $T_C$ (K) | $M_s$ (emu/cm$^3$) | Laser-induced precession |
|--------|---------|-----------|---------------------|--------------------------|
| F010   | 1.5     | 29        | 8.9                 | No                       |
| F008   | 2       | 47        | 11.6                | Yes                      |
| F007   | 2.5     | 60        | 11.5                | Yes                      |
| F002   | 3       | 77        | 16.2                | Yes                      |
| F016   | 3.8     | 96        | 24.7                | Yes                      |
| E101   | 4.5     | 111       | 27.8                | Yes                      |
| F020   | 5.2     | 132       | 33.3                | Yes                      |
| E115   | 7       | 159       | 51.0                | Yes                      |
| E122   | 9       | 179       | 63.7                | Yes                      |
| F056   | 14      | 182       | 78.1                | No                       |

TABLE I: Table summarizing the basic characteristics of selected samples from the series of optimized materials: $x$ is nominal doping, $T_C$ is Curie temperature, $M_s$ is saturated magnetic moment. Last column shows if the impact of ultrafast laser pulse induces a precession of magnetization for the sample temperature of 15 K and the pump pulse fluence of about 30 µJ.cm$^{-2}$.

For simplicity, we show in this Supplementary information only the helicity-independent signals measured in (Ga,Mn)As epilayer with a nominal doping $x = 3\%$, which is the sample that was used for a detailed analysis presented in the main text. The extensive information about the helicity-dependent signals can be found in our previous paper [r11b].

**EXPERIMENTAL DETAILS**

We investigated the laser-pulse induced dynamics of magnetization by a pump-and-probe magneto-optical technique. A schematic diagram of the experimental set-up is shown in Fig. S1. The output of a femtosecond laser is divided into a strong pump pulse and a weak probe pulse that are focused to the same spot on the measured sample. Laser pulses, with the time width of 200 fs and the repetition rate of 82 MHz, were tuned to 1.64 eV, i.e. above the semiconductor band gap, in order to excite magnetization dynamics by photon absorption. The pump pulses were usually circularly polarized (with a helicity controlled by a wave plate)

and the probe pulses were linearly polarized. The measured magneto-optical signals correspond to the probe polarization rotation induced by the pump pulses (see Fig. S1). The experiment was performed close to the normal incidence geometry ($\theta_i = 2°$ and $8°$ for pump and probe pulses, respectively) with a sample placed in a cryostat, which was placed between the poles of an electromagnet. The external magnetic field $H_{ext}$ was applied in the sample plane at an angle $\varphi_H$ with respect to the [100] crystallographic direction in the sample plane (see Fig. S1).

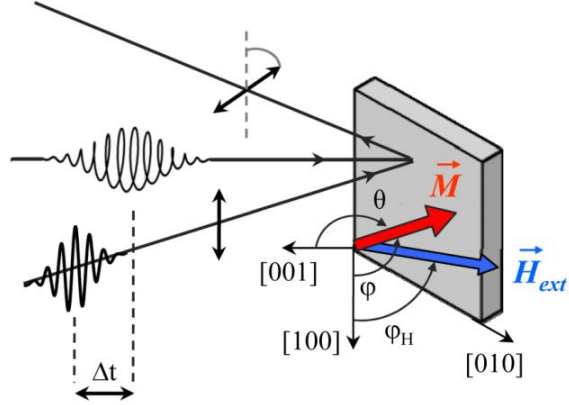

Fig. S1. Schematic diagram of the experimental set-up for a detection of the magnetization precession induced in (Ga,Mn)As by an impact of the circularly polarized femtosecond laser pump pulse. Rotation of the polarization plane of reflected linearly polarized probe pulses is measured as a function of the time delay $\Delta t$ between pump and probe pulses. The orientation of magnetization in the sample is described by the polar angle $\varphi$ and azimuthal angle $\theta$. The external magnetic field $H_{ext}$ is applied in the sample plane at an angle $\varphi_H$.

## DETERMINATION OF THREE-DIMENSIONAL MAGNETIZATION VECTOR TRAJECTORY FROM DYNAMICAL MAGNETO-OPTICAL SIGNALS

In (Ga,Mn)As there are two magneto-optical (MO) effects that are responsible for the measured signal [r12]. In the following we will concentrate on the rotation of the polarization plane $\Delta \beta$ of the reflected linearly polarized light (see Fig. S3(b) for a definition of the probe pulses polarization plane orientation $\beta$) but the same apply also for the change of the light ellipticity (see below). We will also limit the discussion to the case when the incident light is close to the normal incidence. The first of the MO effects is the well-known polar Kerr effect (PKE), which is sometimes called magneto-optical Kerr effect (MOKE), where the rotation of polarization occurs due to the different index of refraction for $\sigma^+$ and $\sigma^-$ circularly polarized light propagating *parallel to the direction of magnetization* - see Fig. S2(a). Consequently, $\Delta \beta$ due to PKE is proportional to the projection of magnetization to the direction of light propagation. We also note that this MO effect is linear in magnetization (i.e., the sign of the

polarization rotation is changed when the direction of magnetization is reversed) and that $\Delta\beta$ is independent on $\beta$ [r12].

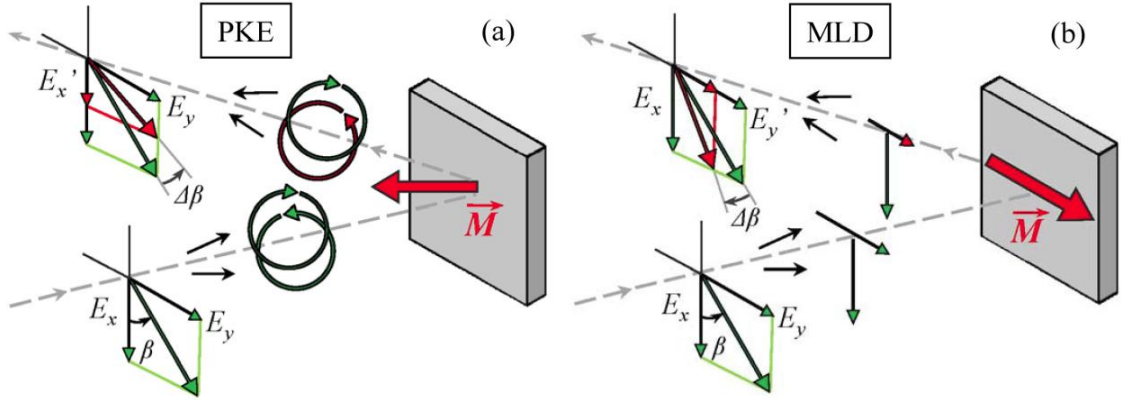

Fig. S2. Schematic illustration of two magneto-optical effects that are responsible for a rotation of the polarization plane $\Delta\beta$ of reflected light at normal incidence in (Ga,Mn)As. (a) Polar Kerr effect (PKE) that is due to the different index of refraction for $\sigma^+$ and $\sigma^-$ circularly polarized light propagating parallel to the direction of magnetization $M$. (b) Magnetic linear dichroism (MLD) that is due to the different absorption (reflection) coefficient for light linearly polarized parallel and perpendicular to $M$ if the light propagates perpendicular to the direction of $M$ [r12].

The second MO effect is the magnetic linear dichroism (MLD), which originates from different absorption (reflection) coefficient for light linearly polarized parallel and perpendicular to $M$, that occurs if the light propagates *perpendicular to the direction of magnetization $M$* - see Fig. S2(b) [r12]. Consequently, $\Delta\beta$ due to MLD is proportional to the projection of magnetization to the direction perpendicular to the direction of light propagation, $\Delta\beta$ is quadratic in magnetization (i.e., the sign of $\Delta\beta$ is not changed when the direction of magnetization is reversed) and it varies as $\sin(2\beta)$ [r12].

For normal incidence of light, the simultaneous presence of two MO effects that are sensitive to the out-of-plane orientation of magnetization (PKE) and the in-plane orientation of magnetization (MLD) enables to perform a full three-dimensional stroboscopic reconstruction of magnetization vector movement from the measured dynamical MO signals [r12]. In particular, different polarization dependences of PKE and MLD enable us to disentangle the pump-induced out-of-plane and in-plane motions of magnetization, respectively [r12]. In equilibrium, the magnetization points to the easy axis direction, which can be characterized by the polar angle $\varphi_0$ and azimuthal angle $\theta_0$ (see Fig. S3(b) for a definition of the coordinate system). The impact of the pump laser pulse induces a precessional motion of magnetization. The magnetization orientation at any time delay $\Delta t$ is given by $\varphi(\Delta t) = \varphi_0 + \delta\varphi(\Delta t)$ and $\theta(\Delta t) = \theta_0 + \delta\theta(\Delta t)$. The time-dependent position of

5

magnetization leads to a time-dependent MO response of the sample that, moreover, depends on the orientation of polarization plane of probe pulses $\beta$ - i.e., the measured dynamical MO signals depend both on $\Delta t$ and $\beta$: $\delta MO(\Delta t, \beta)$. Consequently, it is possible to determine a time evolution of the magnetization vector orientation - without any numerical modeling - directly from the measured MO data [r12]. In particular,

$$\delta\varphi(\Delta t) = [\delta MO(\Delta t, \varphi_0) - \delta MO(\Delta t, \varphi_0 - 90°)]/(4 P^{MLD}), \qquad (1)$$

$$\delta\theta(\Delta t) = -[\delta MO(\Delta t, \varphi_0 - 45°) + \delta MO(\Delta t, \varphi_0 - 135°)]/(2 P^{PKE}). \qquad (2)$$

where $\delta MO(\Delta t, \varphi_0)$ is the pump-induced dynamical MO signal measured by probe pulses with $\beta = \varphi_0$; $\delta MO(\Delta t, \varphi_0 - 90°)$ is the pump-induced dynamical MO signal measured by probe pulses with $\beta = \varphi_0 - 90°$; etc. $P^{PKE}$ and $P^{MLD}$ are MO coefficients that describe the magnitude of PKE and MLD.

We note that for a quantitative evaluation of the magnetization motion from the measured MO signals it is essential to know the magnitude and the sign of the corresponding MO coefficients $P^{PKE}$ and $P^{MLD}$ in the investigated sample [r12]. These coefficients can be directly measured if the magnetization is oriented by a strong external magnetic field to the requested position. For a measurement of $P^{PKE}$, the magnetization has to be oriented out-of-plane (i.e., $\theta = 0°$) and any orientation of light polarization plane $\beta$ can be used. For a measurement of $P^{MLD}$, the magnetization has to be oriented in-plane (i.e., $\theta = 90°$) and $\beta = \varphi_0 - 45°$ has to be used [r12]. In Fig. S3(a) we show the spectral dependence of $P^{PKE}$ and $P^{MLD}$ measured in a (Ga,Mn)As epilayer with nominal doping $x = 3\%$ that is the sample used for a detailed study of the magnetization dynamics at different excitation intensities, as described in the main paper.

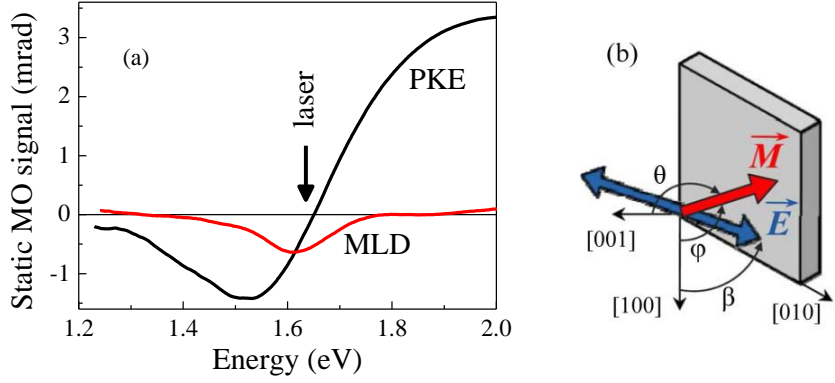

Fig. S3. (a) Spectral dependence of static PKE and MLD in (Ga,Mn)As epilayer with nominal doping $x = 3\%$, the arrow indicate the spectral position of laser pulses used in the time-resolved experiment. (b) The orientation of magnetization $\vec{M}$ in the sample is described by the polar angle $\varphi$ and azimuthal angle $\theta$; the orientation of the polarization plane $\vec{E}$ of probe pulses is described by the angle $\beta$.

In Fig. S4 we show MO hysteresis loops that were measured for the in-plane orientation of the external magnetic field (see the inset in Fig. S4). The observed *M*-shaped hysteresis loop is a signature of the existence of four energetically equivalent easy axis positions of magnetization in a epilayer. Prior to the time-resolved experiment, we prepared the magnetization in a state close to the [010] crystallographic direction (i.e., in an easy axis position labeled "1" in Fig. S4).

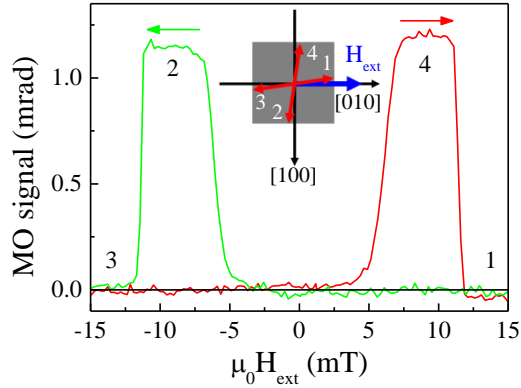

Fig. S4. Magneto-optical hysteresis loops measured in a (Ga,Mn)As epilayer with nominal doping $x = 3\%$ at 15 K; the four energetically equivalent easy axis positions of magnetization are schematically labeled "1" to "4" in the inset.

The impact of a laser pulse on (Ga,Mn)As induces a precession of magnetization by two distinct mechanisms – the helicity-dependent and the helicity-independent [r11b]. To separate these effects, we computed from the dynamical MO signals measured after excitation by $\sigma^+$ and $\sigma^-$ circularly polarized pump pulses the helicity-independent $[(\sigma^+ + \sigma^-)/2]$ and

7

helicity-dependent $[(\sigma^+ - \sigma^-)/2]$ signals. In Fig. S5(a) we show the as-measured data for $\sigma^+$ and $\sigma^-$ excitation. The helicity-independent part of the signal, which is the main focus of this paper, is shown in Fig. S5(b). We note that the helicity-independent signal is identical to that measured with linearly polarized pump pulses [with any orientation of the polarization plane – see Fig. S5(b)].

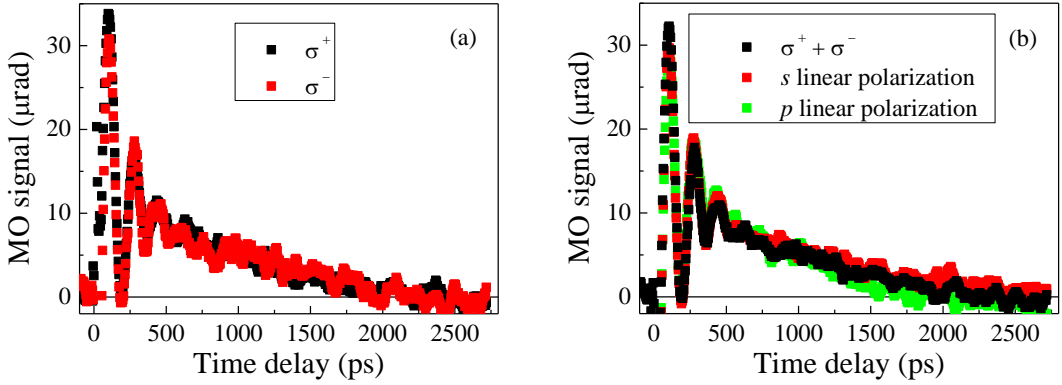

Fig. S5. (a) Dynamical magneto-optical (MO) data measured for circularly ($\sigma^+$, $\sigma^-$) polarized pump pulses. (b) Comparison of the helicity-insensitive $[(\sigma^+ + \sigma^-)/2]$ part of the signals shown in (a) with the MO data measured for linearly polarized ($s$, $p$) pump pulses. Excitation intensity $I_0 = 7$ µJ.cm$^{-2}$, sample temperature $T = 15$ K, probe polarization orientation $\beta = 0°$, no external magnetic field applied.

The dependence of helicity-independent dynamical MO signal on the probe polarization orientation $\beta$ is shown in Fig. S6(a). The measured data can be fitted well by the phenomenological equation

$$\delta MO(t) = A\cos(\omega_{Mn} t + \Delta) e^{-t/\tau_G} + C e^{-t/\tau_p}, \tag{3}$$

where $A$ and $C$ are the amplitudes of the oscillatory and pulse function, respectively, $\omega_{Mn}$ is the ferromagnetic moment precession frequency, $\Delta$ is the phase factor, $\tau_G$ is the Gilbert damping time, and $\tau_p$ is the pulse function decay time. All the measured data in Fig. S6(a) can be fitted well by Eq. (3) with a one set of parameters $\omega_{Mn}$, $\tau_G$ and $\tau_p$. The dependence $A(\beta)$ obtained by this fitting procedure is displayed in Fig. S6(b). The position of the maximum in the dependence $A(\beta)$ at $\beta \approx 100°$ corresponds to the equilibrium position of magnetization in the sample $\varphi_0$ [r12]. We recall that prior to this measurement we prepared the magnetization in the easy axis labeled "1" in Fig. S4, which close to [010] crystallographic direction.

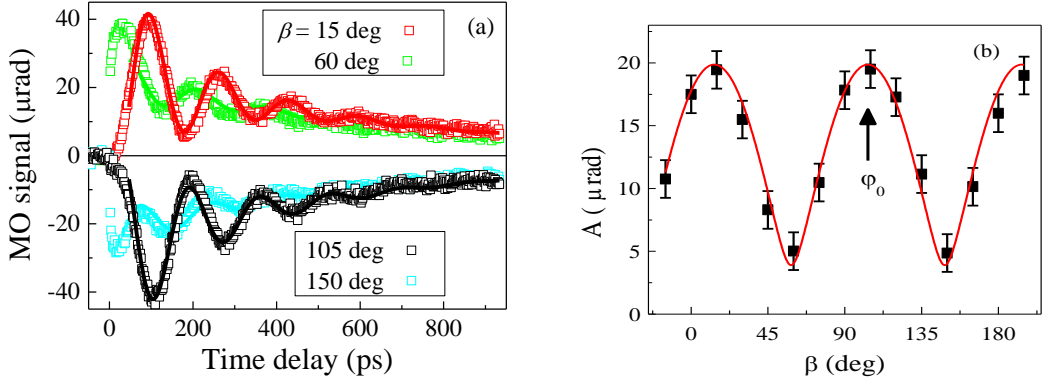

Fig. S6 (a) Dynamics of the helicity-independent MO signal induced by an impact of pump pulse on the sample that was measured by probe pulses with different probe polarization orientations $\beta$ (points). (b) Probe-polarization dependence of the amplitude of the oscillatory part $A$ that was obtained by fitting the measured MO dynamics by Eq. (3) [lines in part (a)] with a fixed values $\omega_{Mn}$ = 33.9 GHz, $\tau_G$ = 165 ps, and $\tau_p$ = 880 ps (points). The line is a fit (see Ref. r12 for details). The vertical arrow depicts the derived easy axis position in the sample. Excitation intensity $I_0$ = 7 μJ.cm$^{-2}$, sample temperature $T$ = 15 K, no external magnetic field applied.

In Fig. S7 we show the dynamics of magnetization that was deduced from the data depicted in Fig. S6(a) using Eqs. (1) and (2) – in part (a) and (b) the time evolutions of $\delta\varphi$ and $\delta\theta$, and the corresponding polar plot are depicted, respectively.

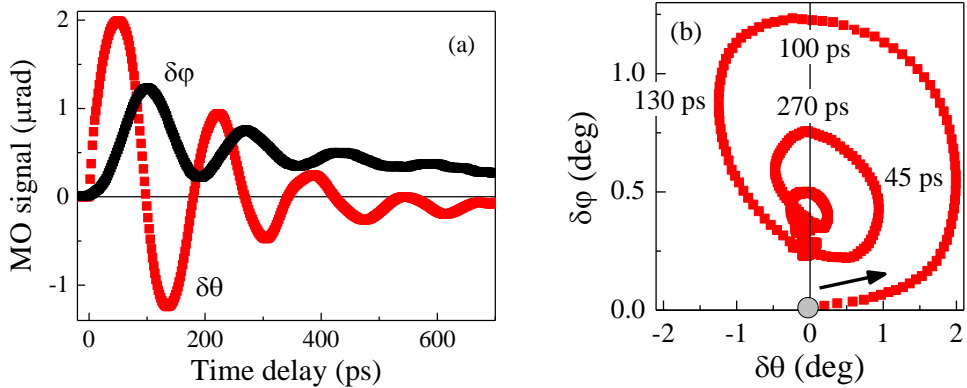

Fig. S7 Magnetization real-space trajectory deduced from the MO data shown in Fig. S6. (a) Time evolution of $\delta\varphi$ and $\delta\theta$. (b) Orientation of magnetization at different times after the impact of the pump pulse; the equilibrium position of the easy axis is depicted by the grey spot.

In principle, the measured dynamical MO signals can contain information not only about the pump-induced magnetization change but also about the pump-induced change of the complex index of refraction (the so-called "optical part" of the MO signal) [r13]. Consequently, the dynamics of both the rotation and ellipticity has to be measured and compared before the obtained magneto-optical signal is attributed to the magnetization dynamics [r13]. In Fig. S8 we show the corresponding data measured at pump intensity $18I_0$ =

9

126 µJ.cm$^{-2}$, which is in the middle of the investigated intensity range (see the main paper). The measured dynamics of the rotation and ellipticity are identical that confirms that the procedure described above can be meaningfully used for a reconstruction of the magnetization movement. In particular, we would like to note that the displayed data which were measured for $\beta = 105° \approx \varphi_0$ provide direct information about the in-plane, $\delta\varphi$, movement of the magnetization [cf. Eq. (1)]. At low excitation intensities, the MO data measured for this value of $\beta$ show a monotonous rise of the (negative) dynamic MO signal for time delays $0 < \Delta t < 50$ ps - see Fig. S6(a) for intensity $I_0 = 7$ µJ.cm$^{-2}$. Taking into account that $P^{MLD} < 0$ in the used spectral range [see Fig. S3(a)], this leads to $\delta\varphi > 0$ – i. e., the magnetization starts its movement towards the [-110] crystallographic direction in the sample (see Fig. S7). On the contrary, at high excitation intensities the measured MO signal has a opposite sign for a time delay $0 < \Delta t < 50$ ps - see Fig. S8 for intensity $18I_0 = 126$ µJ.cm$^{-2}$ - and, therefore, $\delta\varphi < 0$ in the initial stages of the magnetization movements – i. e., the magnetization starts its movement towards the [010] crystallographic direction in the sample (see also Fig. 3 in the main paper).

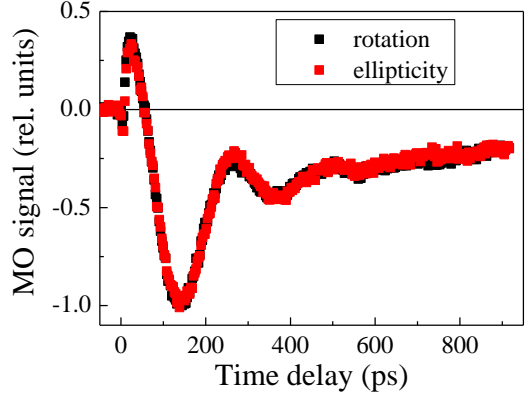

Fig. S8. Comparison of the laser pulse-induced change of the rotation and ellipticity; the data are normalized. Excitation intensity $I = 18I_0 = 126$ µJ.cm$^{-2}$, sample temperature $T = 15$ K, probe polarization orientation $\beta = 105°$, no external magnetic field applied.

# CALIBRATION PROCEDURE USED FOR EVALUATION OF THE PUMP-INDUCED TEMPERATURE AND HOLE CONCENTRATION INCREASE

Absorption of pump laser photons leads to a photo-injection of electrons and holes with identical concentrations. However, because the density of states is considerably larger for holes (due to their larger mass), the pump-induced change of the distribution function is much larger for electrons and, consequently, the measured change of complex index of refraction is dominated by electrons [r14]. Therefore, the dynamics of the reflectivity change $\Delta R/R$ – that is shown in Fig. S9(a) – provide information about the lifetime of photo-injected electrons in the investigated sample. From the measured data we can conclude that the population of photo-injected free electrons decays with a time constant (defined as a decrease to 1/e) of ≈ 15 ps. This rather short lifetime of free electrons is similar to that reported for the low temperature grown GaAs (LT-GaAs), which is generally interpreted as a result of a high concentration of nonradiative recombination centers induced by the low temperature growth mode of the MBE [r15]. The nonradiative recombination of photo-injected electrons is accompanied by an emission of phonons (i.e., by a heating of the sample). Consequently, the measured decay time of $\Delta R/R$ corresponds to a rise time of the laser-induced transient change of the sample temperature $\delta T$. We note that the lifetime of the photo-injected holes can be expected to be longer than that of electrons because (Ga,Mn)As is a heavily *p*-doped material and, therefore, the traps for holes are at least partially saturated by the background holes.

We applied the following analysis for an evaluation of the laser-induced change of the hole concentration $\delta p$. The initial value of $\Delta R/R$ can be regarded as a measure of the relative number of the photo-generated electrons (and thus also of the holes). It is apparent from the inset in Fig. S9(a) that the pump-induced change of the complex index of refraction is linear in a pump-pulse intensity up to $I \approx 25I_0 = 180$ μJ.cm$^{-2}$ after which the initial values of $\Delta R/R$ start to saturate. The thickness of the investigated (Ga,Mn)As film (20 nm) is considerably smaller than the penetration depth of the laser at the corresponding photon energy (≈ 600 nm). Consequently, the absorption of photons leads to a rather constant concentration of photo-injected carriers in (Ga,Mn)As along the growth direction, which can be estimated from the laser spot size on the sample, photon energy, and absorption and reflection coefficients of GaAs. On the other hand, also the carriers photo-injected in the substrate could, in principle, contribute to the measured signals but we have verified that the measured signals in (Ga,Mn)As samples on GaAs substrate are markedly different from that in a bare GaAs

substrate. In Fig. S10 we show the estimated intensity dependence of the initial concentration of the excess holes $\delta p$.

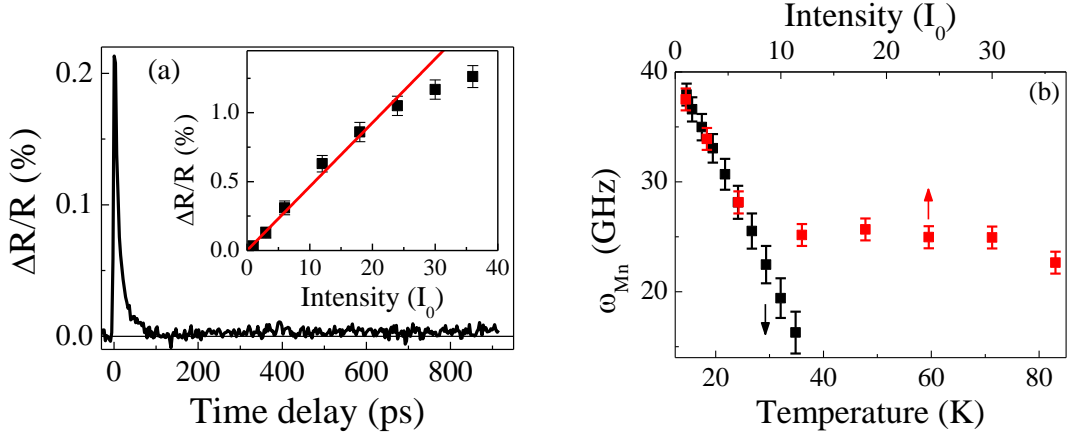

Fig. S9 (a) Dynamics of pump pulse-induced reflectivity change $\Delta R/R$ at intensity $6I_0 = 42$ μJ.cm$^{-2}$; Inset: Intensity dependence of the initial value of the reflectivity change at 15 K (points), line depicts the linear dependence. (b) Frequency of precessing Mn moments measured (without external magnetic field) as a function of base temperature at low excitation intensity $I_0 = 7$ μJ.cm$^{-2}$ and as a function of the laser intensity at low base temperature of 15 K. This figure is re-plotted from the main paper for convenience.

As already described above, the fast nonradiative recombination of photo-injected electrons induces a transient pump pulse-induced increase of the lattice temperature $\delta T$ within tens of picoseconds after the impact of the pump pulse and this temperature increase persists over hundreds of picoseconds. An estimate of $\delta T$ can be obtained from the measured precession frequencies because they correspond to magneto-crystalline anisotropy fields in the sample [r11b], which are rather strongly temperature dependent. In Fig. S9(b) we plot the dependence of the precession frequency on the base sample temperature at low excitation intensity $I_0$ and on the laser intensity at low base temperature of 15 K. From the comparison of these two measurements we infer the magnitude of the transient temperature change $\delta T$ as a function of the laser intensity, which is shown in Fig. S10. (We note that consistent temperature vs. intensity calibration is obtained from the comparison of the intensity dependence of the pump-induced demagnetization and the temperature dependence of the remanent magnetization measured by SQUID.) Remarkably, the temperature increase saturates at a considerably lower pump intensity than the hole concentration increase does that enables us to identify clearly the influence of the latter in the magnetization dynamics (see the main paper).

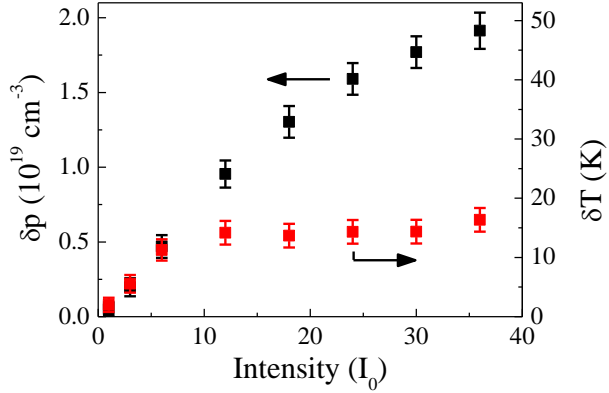

Fig. S10 The pump-pulse induced concentration of the excess holes $\delta p$ and the transient temperature increase $\delta T$ as a function of the laser intensity.

## QUALITATIVE THEORY OF OSOT

Absorption of pump laser pulse with a photon energy of 1.64 eV leads to photo-injection of charge carriers with a rather high excess energy (the low-temperature band gap energy of GaAs is 1.52 eV). Immediately after photoinjection, the carriers have a non-thermal distribution. The Fermi-Dirac distribution is established on a time scale from tens to hundreds of femtoseconds (mainly due to mutual collisions between the carriers) but the resulting characteristic carrier temperature significantly exceeds that of the lattice. The carrier temperature starts to cool down towards the lattice temperature (by an emission of phonons) on a time scale from hundreds of femtoseconds to several picoseconds and, finally, the non-equilibrium carriers recombine. During this process holes relax through the available states in the spin-split valence band of the ferromagnetic semiconductor. Consequently, a change in the occupation of the hole states, as compared to the equilibrium state in dark, leads – due to the spin-orbit coupling – to a non-equilibrium hole spin polarization which is misaligned with the equilibrium orientation of Mn moments. This non-equilibrium photo-hole polarization persists over the timescale of the hole recombination during which it exerts a torque on the Mn local moments via the kinetic-exchange coupling. The exact theoretical treatment of the mutual interplay of all the involved effects is rather complicated and, therefore, derivation of a full quantitative theory of OSOT is a challenging problem. (We note that in the case of OSTT the non-equilibrium spin-density of weakly spin-orbit coupled photo-electrons, producing the OSTT, is directly determined by the external polarizer - i.e., by the intensity, propagation axis, and helicity of the circularly polarized pump laser beam – which make the corresponding theory much simpler [11b]). As a first proxy to the complex theory of the non-equilibrium

photo-hole spin polarization, related to OSOT by Eq. 1 in the main text, we consider the hole density dependent magnetic anisotropy fields (see Eq. 2 in the main text). The gross simplification in this qualitative theory is that we consider OSOT produced by fully equilibrated photo-holes acting on the local moments over the recombination time. Within this approximation, the misalignment of the non-equilibrium hole polarization with the equilibrium orientation of Mn moments has the same physical origin as the dependence of the static magnetic easy-axis orientation on the hole density [r16].

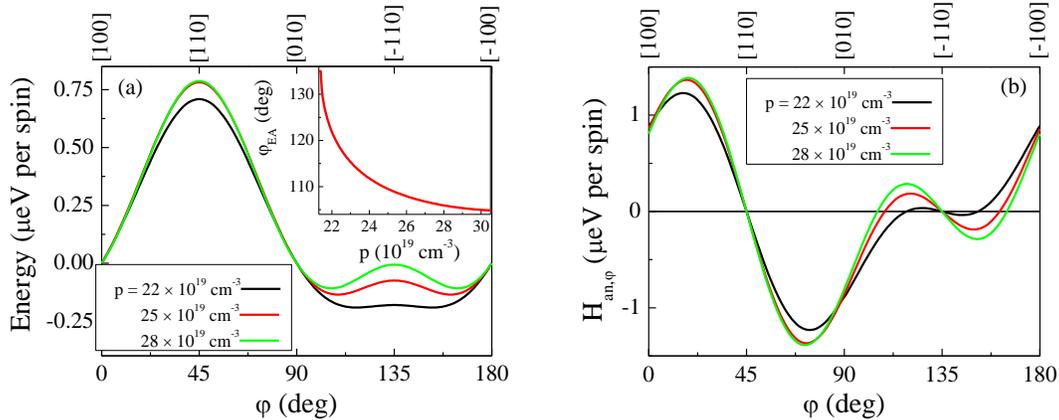

Fig. S11 Computed in-plane angular dependence of the anisotropy energy $E$ (a) and the anisotropy field $H_{an,\varphi} = \delta E/\delta\varphi$ (b) in (Ga,Mn)As with 3% of Mn and strain $\varepsilon_{xy} = 0.01\%$ at 0 K. Inset: Dependence of the easy axis position on the concentration of holes.

In Fig. S11(a) we show the in-plane angular dependence of the anisotropy energy in (Ga,Mn)As, which was computed using the **k·p** kinetic-exchange Hamiltonian [r16]. The position of the energy minimum (i.e., the easy axis position) depends strongly on the concentration of equilibrium holes [see the inset in Fig. S11(a)]. The femtosecond pump-pulse induced change of hole concentration $\delta p$ (from the sample equilibrium hole concentration at dark $p_0$) leads to an abrupt in-plane movement of the energy minimum and, consequently, to a misalignment of the energy minimum and the equilibrium orientation of magnetization **M**. The resulting anisotropy field, which is aiming at having them aligned again, has a zero out-of-plane component $H_{an,\theta} = \delta E/\delta\theta$ and a non-zero in-plane component $H_{an,\varphi} = \delta E/\delta\varphi$ [see Fig. S11(b)]. Consequently, the resulting quasi-instantaneous torque acting on **M** is oriented in the out-of-plane direction (see Eq. (1) and Fig. 1(b) in the main paper) which explains why we observed the fast out-of-plane magnetization tilt for high excitation intensities [see Fig. 3(c) and (d) in the main paper]. On the other hand, at low excitation intensities the pump-induced temperature increase is considerably slower and the resulting temperature-related movement

of the energy minimum (quasi-equilibrium easy axis) is in the opposite direction than in the case of the hole concentration-related movement [see Fig. 4(b) in the main paper]. This explains why we observed at low excitation intensities a much slower rise of the out-of-plane component of the precessing magnetization and towards the opposite direction than in the case of OSOT at high excitation intensity [see Fig. 3(a) and (b) in the main paper]. The simplified analysis therefore provides a physically transparent qualitative description of the experimentally observed OSOT.


\end{document}